\documentstyle[11pt,paspconf,epsf]{article}

\begin{document}

\title{Modeling the Clustering of Objects}

\author{E. Salvador-Sol\'e}
\affil{Department d'Astronomia i Meteorologia, Universitat de Barcelona,
    Barcelone, E--08028 SPAIN}

\begin{abstract}

I review the main steps made so far towards a detailed (semi) analytical
model for the hierarchical clustering of bound virialized objects (i.e.,
haloes) in the gravitational instability scenario. I focus on those models
relying on the spherical collapse approximation which have led to the most
complete description. The work is divided in two parts: a first one dealing
with the mass function of objects and a second one dealing with the growth
times and rates.

\end{abstract}

\keywords{cosmology: theory -- galaxies: clustering -- galaxies: formation}

\section{Introduction}

Observational data of cosmological relevance refer to the characteristics
of the CMB radiation and the clustering and structural properties of bound
virialized (more exactly, relaxed and steady) objects such as Lyman-$\alpha$
clouds, galaxies, and galaxy clusters. To take full benefit of the
information contained in the latter kind of data a good knowledge of how,
when, and where these objects formed and evolved is required. Indeed, this
would allow us not only to correctly interpret the observed properties of
those cosmological objects but also to properly use them to constrain the
correct cosmogony (i.e., the possible Gaussianity of the initial density
field, its power spectrum, and the values of $\Omega$, $\Lambda$, and $H_0$).
Unfortunately, the modeling of the formation and growth of cosmic objects is
not an easy task. Even in the simple and yet most likely scenario, hereafter
assumed, of structure formation via gravitational instability from a
primordial Gaussian random field of density fluctuations with power spectrum
leading to hierarchical clustering no exact model can be build. The reason
for this is the lack of an exact solution for the growth of density
fluctuations in the non-linear regime.

There are only two ways to circumvent this difficulty: the use of numerical
simulations and the construction of (semi) analytical models relying on
approximated collapse dynamics. The former is obviously more exact but it is
not free of problems, either. Numerical simulations are very time-consuming,
which translates into a limited dynamical range and a very reduced region of
the parameter space covered. Moreover, numerical simulations give access to
the yields of the complex processes taking place, but the full understanding
of what is going on is not easy. In contrast, models are less accurate,
sometimes possibly poorly justified, but are more practical and allow a
deeper insight into the physics. In fact, both approaches are complementary:
simulations ultimately justifies the goodness of analytical models while the
latter bring the possibility to confortably explore a wide range of
parameters and allow to better understand the results of the former. There
are in the literature numerous reviews dealing with cosmological simulations.
Here I will focus on the improvements achieved, for the last twenty years, in
the construction of a detailed model for the hierarchical clustering of
objects.

The different models developed so far are of two main kinds. On the one
hand, there are models developed to derive the theoretical mass function of
objects (or haloes). These are brievely reviewed in \S\ 2, the most relevant
ones being discussed in more detail in \S\ 3. On the other hand, there are
models which go further and provide us with typical times and rates of
the clustering process. These latter models are addressed in \S\ 4. For
simplicity, I assume an Einstein-de Sitter ($\Omega=1$, $\Lambda=0$) universe
and comoving units.

\section{Theoretical Mass Functions}

As mentioned, all clustering models are based on some approximation to the
collapse dynamics of density fluctuations. Most of them, in particular the
seminal model by Press \& Schechter (1974; PS), rely on the spherical
collapse model. This is a poor approximation, in general, to the real
collapse. Yet, the PS mass function gives very good fits to $N$-body
simulations (Nolthenius \& White 1987; Efstathiou et al. 1988; Efstathiou \&
Rees 1988; Carlberg \& Couchman 1989; White et al. 1993; Bahcall \& Cen 1993;
Lacey \& Cole 1994). The reason is likely that massive objects, those
intended to be described, arise from high amplitude peaks (density maxima) of
the initial density field and the collapse of matter around such peaks is
particularly well described by the spherical model (Bernardeau 1994).
$N$-body simulations seem to show that there is no good correspondence
between peaks and objects (van de Weygaert \& Babul 1994; Katz, Quinn, \&
Gelb 1993). But this is likely due to a variety of effects, namely the
nesting of peaks on different scales, the use of an unappropriated window, or
the inclusion of density constrasts and masses which do not correspond to the
collapse times and filtering scales analyzed (see below).

Actually, the PS mass function and new more or less sophisticated versions of
it (Cole \& Kaiser 1989; Bond et al. 1991, BCEK; Blanchard, Valls-Gabaud, \&
Mamon 1994; Jedamzik 1995; Yano, Nagashima, \& Gouda 1995) do not explicitly
deal with peaks as seeds of bound objects. But a parallel set of models has
also been developed within the peak theory framework (Colafrancesco, Lucchin
\& Matarrese 1989; Bond 1989; Peacock \& Heavens 1990; Apple \& Jones 1990;
Manrique \& Salvador-Sol\'e 1995) reaching similar results.

In the context of models relying on the spherical approximation, we must also
mention the model constructed by Cavaliere \& Menci (1994) using the theory
of Cayley trees with disorder. This is a more general formalism which
recovers, as two extreme limits, the diffusion equation describing, as shown
by BCEK, the clustering of objects \`a la PS, and the Smoluchowski kinetic
equation describing the aggregation of objects moving inside a relaxed
system. Indeed, this formalism is intended to derive the mass function of
objects accounting for the fact that may survive and evolve inside larger
scale objects (for example, galaxies in clusters). This mass function is
different from that intended to be derived in all previous models; these only
consider relaxed haloes which are not embedded within any larger scale
relaxed system. Here we will focus on the latter most usual viewpoint.

There are also a few models based on other dynamical approximations. Monaco
(1995) has followed the PS approach but using the ellipsoidal collapse
approximation. Bond \& Myers (1993a, 1993b) have considered this latter
approximation in the framework of the peak theory. Finally, Doroshkevich \&
Kotok (1990) and Vergassola et al. (1994) have used the adhesion model.
In principle, these are better approximations to the true collapse than the
simple spherical model. However, in the case of the adhesion approximation,
the mathematical calculations are very complicated and one can only infer
approximate analytical solutions for the cases of pure 1-D, 2-D, or 3-D
collapsed structures (see Doroshkevich \& Kotok 1990). For the real composite
case, one can only obtain the asymptotical behavior (Vergassola et al. 1994).
Concerning the mass function obtained by Monaco (1995), it is not clear why
it does not recover the PS solution at the large mass end where the spherical
approximation should be essentially correct. In fact, Bond \& Myers (1993a,
1993b) find, in contrast, that the spherical collapse is a good approximation
for very massive objects, indeed. The only drawback of the very accurate
approach followed by these latter authors (the so-called ``peak-patch''
formalism) is that it involves complicated calculations including
Monte-Carlo simulations which makes it less handy than usual (semi)
analytical models.

\section{Models based on the Spherical Collapse Approximation}

\subsection{The PS Mass Function}

According to the spherical collapse model (Gunn \& Gott 1972), the collapse
time $t$ for a shell of radius $R$ around the center of a spherically
symmetric, outwards decreasing (to avoid shell crossing until $\sim t$)
linear density fluctuation partaking of the general Hubble expansion at $t_i$
only depends on the mean density contrast $\delta$ (the density fluctuation
normalized to the mean density of the universe) inside it through the
relation $\delta(t) =\delta_{c0}\,a(t_i)/a(t)$, with $a(t)$ the cosmic
expansion factor and $\delta_{c0}$ a constant equal to $3/20\,(12\pi)^{2/3}
\approx 1.69$. The collapse of that shell represents, of course, the
appearance, at $t$, of a relaxed object of mass equal to (to 0th order in
$\delta$) $4\pi/3\,\rho\,R^3$, with $\rho$ the mean density of the universe.

Inspired of this simple model, PS assumed that any point in the initial
(linear and Gaussian distributed) density field smoothed with a top-hat
filter of scale $R$ with density contrast {\it above} the
overdensity $\delta_c$ collects matter so to reach, at $t$ related to
$\delta_c$ through the expression above, a mass {\it larger than\/}
$M(R)=4\pi/3\,\rho\,R^3$. Consequently, by differentiating over $M$ the
volume fraction occupied by such points,
\begin{equation}
f(\ge\delta_c,R)\,={1\over2}\,{\rm erfc}\bigg[{\delta_c\over \sqrt2
\,\sigma_0(R)}\bigg],\label{e2}
\end{equation}
with $\sigma_0(R)$ the rms density contrast on scale $R$, one should obtain
the volume fraction contributing at $t$ with objects of mass $M$ to $M+dM$,
and by dividing it by $M/\rho$ the number density of such objects
\begin{equation}
N(M,t)\,dM\,=\,2\,{\rho\over M}\,\biggl|{\partial
f(\ge\delta_c,R)\over \partial R}\biggr|\, {dR\over dM}\,dM.\label{e1}
\end{equation}
It is worthwhile mentioning that, in the case of power-law power spectra,
there should be no privileged time or scale (in an Einstein-de Sitter
universe as assumed here). The PS mass function recovers this expected
behavior. The number of objects in a volume $M_*/\rho$, with $M_*$
corresponding to a scale defined through any arbitrary fixed value of
$\sigma_0(R_*)$, with mass $M/M_*$ in an infinitesimal range, as well as the
volume (or mass) fraction subtended by objects of scaled mass $M/M_*$ in an
infinitesimal range are time invariant, indeed.

But the growth of density fluctuations can deviate from the spherical
collapse in leaving the linear regime. Hence, one should check whether small
changes in those aspects the most strongly connected with the spherical
approximation are suitable. In particular, other filters than the top-hat
one, and other values of constant $\delta_{c0}$ or of the proportionality
factor $q^3$ between the mass and $\rho$ times the natural volume of the
filter should be investigated. (We must remark that there is degeneracy
between the latter two constants, so there is just one degree of freedom for
any given filter.) Yet, Lacey \& Cole (1994) have recently shown that a very
satisfactory fit to $N$-body data can be obtained for masses in the relevant
range with a top-hat filter and $\delta_{c0}$ close to the standard value
(for $q=1$).

A more serious problem, apart from the unnatural seeds of bound objects
assumed, concerns the unjustified factor two in the right-hand member of
equation (\ref{e1}). This must be introduced for the final mass function to
be correctly normalized, that is, for the integral of $M$ times the mass
function to be equal to the mean density of the universe. (Every particle in
the universe is at any time $t$ within some virialized object with
appropriate mass.) On the other hand, the overcounting of objects actually
swallowed by previously collapsed ones and the neglected contribution to
the mass of objects of low density regions enclosed within high density
ones (which might explain the fudge factor 2) are not accounted for. To
analyze the effects of such cloud-in-cloud configurations Cole \& Kaiser
(1989) have devised a practical numerical method, called the ``block model''.
After decomposing (through a series of cuts in two pieces) a large cuboidal
volume in very small cuboidal blocks with different overdensities (assigned
at each level, through Monte Carlo, according to the Gaussian distribution
corresponding at that scale) one can follow their detailed merger trees free
of the cloud-in-cloud problem under the same clustering assumptions as in the
PS approach (except for the rather unnatural geometry of the cuboidal
filter).

\subsection{The Excursion Set Formalism}

A more satisfactory solution to these latter problems, in the sense that
not attached to any particular realization (using a spherical filter) and
leading to a fully analytical solution, was provided by BCEK by means of the
powerful ``excursion set'' formalism. When the filter size is increased, the
density contrast at a fixed point can diminish or increase depending on
whether the original cloud is embedded in a higher density contrast one or
not. So the random walk followed by this point in the $\delta$ vs. $R$
diagram will inform us on the nesting of clouds centered on that point. In
particular, the mass of the only object which must be counted at $t$ attached
to a fixed point is given by the largest scale $R$ for which the $\delta_c$
line is upcrossed.

The mathematical description of such random walks is hard to achieve in
general. However, for the sharp k-space filter, the volumes subtended by
different scales $R$ are uncorrelated. Consequently, the random walk followed
by $\delta(R)$ is then purely Brownian with variance $\sigma_0^2=(R)\equiv S$
and the equation describing the number density $Q(S,\delta)$ of trajectories
found at $(S,\delta)$ which start at $(S_0,\delta_0)$ is the simple diffusion
equation
\begin{equation}
{\partial Q\over \partial S}={1\over2}\,{\partial^2 Q\over\partial
\delta^2}.\label{diff}
\end{equation}

Therefore, the volume fraction in objects with mass in the range
$M$ to $M+dM$, equal to the probability that a trajectory starting at
$(S_0=0,\delta_0=0)$ (corresponding to the limit for $R=\infty$ of the
smoothed density contrast attached to any fixed point) upcrosses for the
first time $\delta_c$ in the corresponding range of $S$, is simply given by
the reduction, in that range, in the number density of trajectories
surviving below $\delta_c$
\begin{equation}
f(\delta_c,S)\,dS= \biggl[-{\partial \over \partial
S}\int_{-\infty}^{\delta_c} Q(\delta,S) d\delta\biggr]\,dS,\label{frac}
\end{equation}
with
\begin{equation}
Q(\delta,S)={1\over\sqrt{2\pi}}\biggl\{\exp\biggl(-{\delta_c^2\over
2S}\biggr) -\exp\biggl[{(\delta-\delta_c)^2\over 2S}\biggr]\biggr\}
\end{equation}
the solution of the diffusion equation (\ref{e1}) with absorbing barrier at
$\delta_c$. Interestingly enough, the solution one gets (after changing to
variable $M$) is just the PS mass function with the correct normalization
factor 2. But, why the sharp k-space filter?

\subsection{An Improved Correction for the Cloud-in-Cloud}

Moreover, the previous formalism only corrects for nested configurations
which are well centered on each fixed point; off-center nested configurations
are not accounted for. To better correct for the cloud-in-cloud one must
abandon the excursion set formalism. (This considers the evolution in the
$\delta$ vs. $R$ diagram of each fixed point separately, that is, it cannot
see any correlation of the density field among different points.) Jedamzik
(1995) proposed to directly apply the PS prescription, equation (\ref{e1}) to
the volume fraction (\ref{e2}) uncorrected for any nesting, denoted here by
subindex $PS$, minus the volume fraction in clouds nested within any larger
scale cloud with $\delta_c$
\begin{equation}
f(\ge \delta_c,R)=f_{PS}(\ge \delta_c,R)-{1\over\rho}\int_R^\infty
M(R')\,N(R',\delta_c)\,P(\ge\delta_c,R|\delta_c,R')\,dR.\label{e3}
\end{equation}
In writing equation (\ref{e3}) we have taken into account the remarks by
Yano, Nagashima, \& Gouda (1995) on its correct expression.
$P(\ge\delta_c,R|\delta_c,R')$ is the probability that a cloud of size $R$
with $\delta\ge \delta_c$ is located on a background with $\delta=\delta_c$
on scale $R'$, while $M(R')\,N(R',\delta_c)\,dR'/\rho$ approximately gives
the probability that such a background is found inside a non-nested cloud
with $\delta_c$ on scale in the range $R'$ to $R'+dR'$. The probability $P$
can be easily calculated in the case of sharp k-space filter, since the
probability of finding two values of $\delta$ on different scales at a given
point is then simply the product of finding each of them separately.

$N(R,\delta_c)\,dR$ is the unknown scale function, i.e., the mass function
previous to the change to variable $M$, that we want to determine. Therefore,
by applying the PS prescription to equation (e3) one is led to a Volterra
type integral equation of the second kind for $N(M,t)$ which can be readily
solved through the standard iterative method from the initial approximated
solution given by the PS mass function. (This is equivalent to the practical
algorithm proposed by Jedamzik to solve its equation.)

\subsection{The PS Approach Extended to the Peak Model}

But peaks are better motivated seeds of objects than the fuzzy regions
considered in all previous models. Also inspired of the spherical collapse
model, the ``peak model'' ansatz states that objects at a time $t$ emerge
from peaks with density contrast equal to a fixed linear overdensity
$\delta_c$ in the smoothed, on any scale $R$, density field at the arbitrary
initial time $t_i$. The critical overdensity is assumed to be a monotonous
decreasing function of $t$, while the mass $M$ of objects can also be assumed
(the consistency of this guess is to be confirmed a posteriori) a monotonous
increasing function of $R$.

The PS prescription (equation \lbrack\ref{e2}\rbrack) is therefore achieved,
in
this framework, by simply taking (Colafrancesco, Lucchin, \& Matarrese 1989;
Peacock \& Heavens 1990)
\begin{equation}
f(\ge \delta_c,R)= n_{pk}(\delta_c,R)\,{M_{pk}(\delta_c,R)\over\rho},
\label{e4}
\end{equation}
where $n_{pk}(\delta_c,R)$ is the number density of peaks with
$\delta\ge\delta_c$ in the density field smoothed on scale $R$, calculated by
Bardeen et al. (1986; BBKS), and $M_{pk}(\delta_c,R)$ is the average mass of
their respective collapsing clouds, i.e., of the objects giving them rise.
Note that since peaks in $n_{pk}(\delta_c,R)$ do not have, in general,
$\delta=\delta_c$, the average mass of their collapsing clouds,
$M_{pk}(\delta_c,R)$, will differ from $M(R)$. The above mentioned problems
with the normalization of the PS mass function and the cloud-in-cloud are
reflected in the different expressions for $M_{pk}(\delta_c,R)$ found in the
literature.

But there is a more serious problem. In applying equation (\ref{e1}) to the
volume fraction (\ref{e4}) it has been implicitly assumed that: 1) the total
mass in collapsing clouds associated with peaks (with $\delta>0$) is
conserved with varying scale, and 2) the density contrast of peaks is a
decreasing function of scale. This guarantees, indeed, that the variation
along $dR$ of the mass associated with peaks above $\delta_c$ is just that
associated with peaks crossing $\delta_c$ in that infinitesimal range of
scales. Both points seem to follow from the peak model ansatz, but they
actually do not. As shown below, point 2 crucially depends on the shape of
the filter used, while mergers invalidate point 1 in any event.

\subsection{An Extension to Peaks Inspired of the Excursion Set Formalism}

As pointed out by Bond (1988), the only reliable strategy to derive the mass
function in the peak model framework is therefore to directly count the
density of peaks with density contrast upcrossing $\delta_c$ in an
infinitesimal range of scale, $N_{pk}(R,\delta_c)\,dR$, then correct for
the cloud-in-cloud, and finally transform to the mass function of objects
at $t$, $N(M,t)\,dM$, through the appropriate $M(R)$ and $\delta_c(t)$
relations.

Taking into account that for a Gaussian filter we have
\begin{equation}
{d\delta\over dR}=R\,\nabla^2\delta, \label{e8}
\end{equation}
Bond (1989) and Appel \& Jones (1990) derived the wanted scale function of
peaks by computing, \`a la BBKS, the density of peaks at $R$ with the
extra constraint that they cross $\delta_c$ between $R$ and $R+dR$,
\begin{equation}
\delta_c <\delta\leq\,\delta_c-\nabla^2\delta\,R\, dR.
\end{equation}
This leads to
\begin{equation}
N_{pk}(R,\delta_c)\,dR\,=\,{dn_{pk}(\nu,R)\over
d\nu}\bigg|_{\nu=\delta_c/\sigma_0}\,{\sigma
_2(R)\over\sigma_0(R)}\,<x>\,R\,dR,\label{e10}
\end{equation}
where $n_{pk}(\nu,R)$ is the density appearing in equation (\ref{e4}), with
$\nu\equiv \delta/\sigma_0(R)$, and $<x>$ is an average of $-\nabla
\delta/\sigma_2(R)$, with $\sigma_2(R)$ the second order spectral moment,
given in Manrique \& Salvador-Sol\'e (1995a; MSa).

To correct this scale function for the cloud-in-cloud Bond (1989) used
the approximate exclusion factor
\begin{equation}
F(R,\delta_c)=\exp\biggl[-\int_R^\infty dR'\,{M(R')\over\rho}\,
N_{pk}(R',\delta_c)\biggr]
\end{equation}
obtained from the excursion set formalism. Note that this coincides with the
Poisson probability that, in a volume typically harboring one peak on scale
$R$, there is no such peak located in the volume fraction independently
subtended by collapsing clouds associated with larger scale peaks.

But what $M(R)$ and $\delta_c(t)$ relations must we take to transform
this corrected scale function to the wanted mass function, and why should we
use the Gaussian filter? And what is worse, the previous derivation
implicitly assumes that the spatial location of peaks does not change in
varying the filtering scale which is obviously not true in general.

\subsection{The Confluent System Formalism}

To account for this variation MSa have developed a new formalism, the
``confluent system of peak trajectories'', able to follow the filtering
evolution of peaks despite their spatial shift.

To guarantee that one peak on scale $R+\Delta R$, with $\Delta R$ arbitrarily
small, traces the same accreting object as another peak on scale $R$ at the
times corresponding to their respective density contrasts, the separation
between both points must be, at most, of the order of $\Delta R$. In this
manner, the collapsing cloud associated with the peak on scale $R+\Delta R$
will essentially include the whole cloud associated with the peak on scale
$R$. Furthermore, this proximity condition is not only necessary, but also
sufficient: as readily seen from the Taylor series expansion of the density
gradient around a density maximum, there cannot be more than one peak on
scale $R+\Delta R$ in the close neighborhood of any peak on scale $R$. This
identification allows one to draw a $\delta$ vs. $R$ diagram similar to
the excursion set one but for the fact that each trajectory $\delta(R)$ is
now attached to one individual accreting object, i.e., to {\it the changing
peaks tracing it\/} in the filtering process, instead of to one fixed point.
It is shown that the total derivative $d\delta/dR$ of a peak trajectory in
this diagram coincides with the partial derivative $\partial_R \delta$ of the
current peak. Moreover, for the mass of accreting objects to increase with
time, $\delta$ must decrease with increasing $R$ along any peak trajectory,
which is only satisfied {\it for a Gaussian filter}.

The density of peak trajectories upcrossing the $\delta_c$ line in an
infinitesimal range of scales is equal to the density of peaks on scale $R$
with $\delta\ge\delta_c$ {\it evolving} into peaks with $\delta\le\delta_c$
on scale $R+\Delta R$. Given the mandatory Gaussian filter and the form of
the total derivative of $\delta$ over $R$ along a peak trajectory one is just
led to equations ({e8})--({e10}). The important point is that this derivation
is now fully justified.

Moreover, to correct for the cloud-in-cloud MSa followed the more accurate
approach pointed out by Jedamzik (1994). The result is the Volterra integral
equation of the second kind
\begin{equation}
N(R,\delta_c)=N_{pk}(R,\delta_c)-{1\over \rho}\,\int_R^\infty
dR'\,M(R')\,N(R',\delta_c)\,
N_{pk}(R,\delta_c|R',\delta_c).\label{e11}
\end{equation}
In equation (\ref{e11}) $N_{pk}(R,\delta_c|R',\delta_c)\,dR$ is the
conditional density of peaks with $\delta_c$ on scales $R$ to $R+dR$ given
that they have density $\delta_c$ on scale $R'$, which can be written in
terms of the analog density per infinitesimal density contrast calculated by
BBKS in a similar way as equation (\ref{e10}), and
$\rho^{-1}\,M(R')\,N(R',\delta_c) \,dR'$ gives the approximate probability to
find such a point inside the collapsing cloud associated with a non-nested
peak with $\delta_c$ on some scale in the range $R'$ to $R'+dR'$.

Now, if the density field is endowed with a power-law power spectrum the
scale function must be self-similar. Likewise, the mass fraction in objects
with scaled mass $M/M_*$ in an infinitesimal range, as well as the number of
peaks inside the volume $M_*/\rho$ on scales $R/R_*$ in an infinitesimal
range must be invariant. But this is only satisfied provided
$M(R)=\rho\,(2\pi)^{3/2}\,[q\,R]^3$, with $(2\pi)^{3/2}\,R^3$ the natural
volume associated with the Gaussian window and $q$ an arbitrary constant. On
the other hand, the mass function at $t$ is independent of the arbitrary
initial time $t_i$ provided only $\delta_c(t)=\delta_{c0}\,a(t_i)/a(t)$ with
$\delta_{c0}$ an arbitrary constant. (In contrast with the PS case, there is
no degeneracy now between constants $q$ and $\delta_{c0}$.) With these
relations the scale function (\ref{e11}) leads to the wanted mass function
which turns out to be correctly normalized for whatever values of $q$ and
$\delta_{c0}$ governing the exact collapse dynamics. A good fit can be
obtained to the PS mass function at any time $t$ for appropriate values of
these parameters. For non-power-law spectra, the previous $M(R)$ and
$\delta_c(t)$ relations are shown to also approximately hold. In this case,
however, there is one unique value of $q$ yielding the correct normalization
for whatever value of $\delta_{c0}$. Nonetheless, a good fit can also be
obtained to the corresponding PS mass function at any time for an appropriate
value this parameter.

\section{Growth Rates and Times}

Richstone, Loeb, \& Turner (1992) proposed the time derivative of the PS
mass function as an estimate of the formation rate of objects of mass $M$ at
a given epoch. However, this is a very crude estimate since that quantity is
actually equal to the rate at which objects reach mass $M$ {\it minus the
rate at which they leave this state}, both terms having comparable values.

\subsection{The Excursion Set Formalism}

Following the PS original approach Bower (1991) derived the conditional
mass function of objects of mass at some epoch subject to being part of
another object with a given larger mass at a later time. This was
subsequently achieved by BCEK from the excursion set formalism. To do it one
must simply compute the volume fraction in objects with $S$ in an
infinitesimal range (\ref{frac}) given by the solution of the diffusion
equation (\ref{diff}) with barrier $\delta_c$ now with initial condition
($S_0=S',\delta_0=\delta_c'$) corresponding to the more massive object at the
later epoch, instead of (0,0),
\begin{equation}
f(S,\delta_c|S',\delta_c')\,dS=\biggl\{{\delta_c-\delta_c'\over
\sqrt{2\pi}\,(S-S')^{3/2}}\,\exp\biggl[{(\delta_c-\delta_c')^2\over
2\,(S-S')}\biggr]\biggr\}\,dS,
\end{equation}
and proceed in the usual manner.

The resulting conditional mass function $N(M,t|M',t')\,dM$ was used by Lacey
\& Cole (1993; LC; see also Kauffmann \& White 1993) to infer the
instantaneous merger rate of objects of mass $M$ at $t$ into objects of mass
$M'$ to $M'+dM'$
\begin{equation}
r^m(M\rightarrow M',t)\,dM'= \lim_{\Delta t\rightarrow 0}\,{1\over\Delta
t}\, {N(M,t|M',t+\Delta t)\,N(M',t+\Delta t)dM'\over N(M,t)}.
\end{equation}

This clustering model has been shown by Lacey \& Cole (1994) to be in very
good agreement with $N$-body simulations. However, the PS approach is not
fully satisfactory (see \S\ 3). On the other hand, accretion does not play
any role in this model; one can only follow the instantaneous mass increase
of objects, event which is generically called merger. As a consequence, there
is no specific event marking the beginning or the end of any entity, hence,
properly justifying the words formation or destruction of objects. This
is the reason why the age and survival time of any object must be defined in
terms of the relative variation (say, by a factor 2) in mass along the series
of objects with embedded mass connecting with it.

\subsection{The Confluent System Formalism}

The model based on the confluent system formalism is better justified (peaks
are the seeds of objects and simple consistency arguments
unambiguously fix the filter and the $M(R)$ and $\delta_c(t)$ relations to be
used) and makes the effective distinction between merger and accretion.

When an object evolves by accretion (tracing a continuous curve $\delta(R)$
in the $\delta$ vs. $R$ diagram) the volume $M/\rho$ of the collapsing cloud
associated with the corresponding evolving peak increases. This makes
smaller scale peaks to become nested within it. Their $\delta(R)$ curves
experience then a discontinuity in $R$ at a fixed $\delta$ which can be
naturally interpreted as a merger. The net density of peaks with $\delta$ on
scales $R$ to $R+dR$ becoming nested in non-nested peaks with
$\delta-d\delta$ on scales $R'$ to $R'+dR'$, ${\bf N}^d(R\rightarrow
R',\delta)\,dR\,dR'\,d\delta$, can then be accurately calculated (Manrique \&
Salvador-Sol\'e 1995b; MSb). The instantaneous (true) merger or destruction
rate at $t$ for objects of mass $M$ per specific infinitesimal range of mass
$M'$ ($M<M'$) of the resulting objects is, therefore,
\begin{equation}
r^{d}(M\rightarrow M',t)={{\bf N}^d(R\rightarrow R',\delta_c)\over
N(R,\delta_c)}\,\,{dR'\over dM'}\,\biggl|{d\delta_c\over
dt}\biggr|.\label{e20}
\end{equation}
Note that this merger rate is different from that obtained by LC because,
in the latter, captures of infinitesimal objects are included while, in the
former, they are not.

In addition, objects forming in the interval of time $dt$ from the merger of
similarly massive objects are traced by peaks appearing (there is no previous
peak to be identified with) in the corresponding range of density contrasts
$-d\delta$ without being nested. The net density of non-nested peaks
appearing between $\delta$ and $\delta-d\delta$, ${\bf N}^f(R,\delta)\,dR
\,d\delta$, can also be calculated (MSb). This leads to the instantaneous
formation rate at $t$ of objects of mass $M$
\begin{equation}
r^{f}(M,t)= {{\bf N}^f(R,\delta_c)\over N(R,\delta_c)} \,
\biggl|{d\delta_c\over dt}\biggr|.
\end{equation}

Finally, the instantaneous mass accretion rate of objects of mass $M$
follows from the instantaneous scale increase rate of the corresponding peaks
as they evolve along continuous trajectories in the $\delta$ vs. $R$ diagram.
Form equation (\ref{e8}) we have $dR/d\delta =[-x\,R\,\sigma_2(R)]^{-1}$.
Averaging over the scaled Laplacian of each peak $x$ leads to (MSb)
\begin{equation}
r^a_{mass}(M,t)= {1\over
<x>\,\sigma_2\,R}\,\,{dM\over dR} \,\biggl|{d\delta_c\over dt}\biggr|.
\end{equation}

On the other hand, the density $N_{sur}(t)\,dM$ of objects surviving (i.e.,
having not merged, just accreted) until the time $t$ from a typical
population of objects with masses in the range $M_0$ to $M_0+dM$ at $t_0<t$
is given by the solution, for the initial condition
$N_{sur}(t_0)=N(M_0,t_0)$, of the differential equation
\begin{equation}
{d N_{sur}\over dt}=- r^{d}[M(t),t]\,N_{sur}(t)\label{e22}
\end{equation}
with $r^{d}[M(t),t]$ the integral over $M'$ of the specific merger rate
(\ref{e20}). Hereafter, $M(t)$ is the typical mass at $t$ of such accreting
objects, approximately given by the solution of
\begin{equation}
{dM\over dt}=r^a_{mass}[M(t),t],\label{e25}
\end{equation}
with $M(t_0)=M_0$. Hence, by defining the typical survival time,
$t_{sur}(M_0,t_0)$, of objects with masses $M_0$ to $M_0+dM$ at $t_0$ as the
interval of time since $t_0$ after which their density is reduced (owing to
mergers) by a factor $e$, we are led to the equality $t_{sur}=t_d-t_0$,
where the destruction time $t_d(M_0,t_0)$ is given by the implicit equation
$N_{sur}(t_d)={\rm e}^{-1}\,N(M_0,t_0)$.

Likewise, the density $N_{pre}(t)\,dM$ of objects at $t_0$ that already
existed (i.e., they have just accreted matter since) at a time $t<t_0$ is
given by the solution of
\begin{equation}
{d N_{pre}\over dt}=
r^{f}[M(t),t]\,N[M(t),t]-r^d[M(t),t]\,N_{pre}(t),\label{e23}
\end{equation}
with $N_{pre}(t_0)=N(M_0,t_0)$. Thus, by defining the typical age
$t_{age}(M_0,t_0)$ of objects with masses between $M_0$ and $M_0+dM$ at $t_0$
as the interval of time until $t_0$ before which their density (owing to
their progressive formation and possible disappearance) was a factor $e$
smaller, we are led to the equality $t_{age}=t_0-t_f$, where the formation
time $t_f(M_0,t_0)$ is given by the solution of the implicit equation
$N_{pre}(t_f)={\rm e}^{-1}\,N(M_0,t_0)$.

\begin{figure}[hbtp]
\centering
\centerline{\epsfxsize= 18cm\epsfysize=7cm\epsfbox[50 200 550 400]
{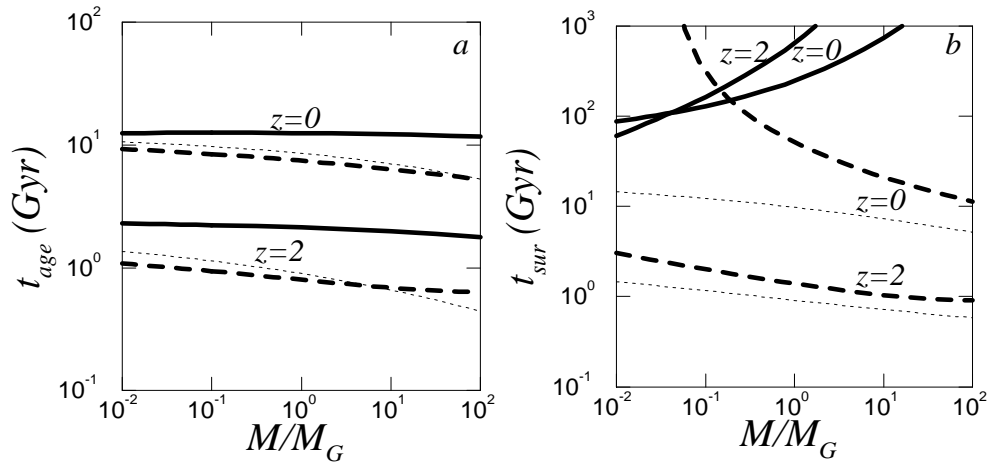}}
\caption{
Age (a) and survival time (b) for objects in the CDM cosmology with
$\sigma_8=$1.5 ($M_G = 10^{12}\,M_{\odot}$). Results obtained by MSb
with the half an double-mass-accretion times in dashed lines (thick)
and LC (thin).} \label{fig-1} \end{figure}

One can also define similar times as those adopted by LC as an estimate of
the age and survival time of obejcts. These are called half-mass-accretion
time and double-mass-accretion and are defined as the interval of time spent
since the mass of an object was typically half its current value and the
interval of time required by an object to typically double its mass,
respectively. (The only difference from the analog times defined by LC is
that, since the new times refer to the typical mass evolution of {\it given
objects\/}, they only involve accretion.) These time estimates can be readily
obtained from equation (\ref{e25}). In Figure~\ref{fig-1} we plot for
comparison the three sets of characteristic times for objects of different
masses at two different epochs.

\end{document}